\documentstyle[12pt]{article}
\def\be{\begin{equation}}
\def\bd{\begin{displaymath}}
\def\ee{\end{equation}}
\def\ed{\end{displaymath}}
\begin{document}

\hfill SNUTP-92-105

\hfill December 1992

\vskip 1.5cm
\begin{center}
{\bf {\huge Neutrino masses from discrete gauge symmetries}}
\vskip 2cm
\Large{Jihn E. Kim and B-{\AA} Lindholm\\
\vskip .8cm
Center for Theoretical Physics\\
Seoul National University\\
Seoul 151-742}\\
Korea
\end{center}
\vskip 3.5cm
{\center Abstract\\}
\vskip .5cm
We investigate a model with an extra $Z_{2}$ gauge symmetry in
the Standard Model.
The symmetry gives a structure to the mass matrix for the neutrinos.
With two extra Higgs singlets and two extra singlet right-handed neutrinos
we can build a model that fits the requirements of the MSW-solution of the
Solar neutrino problem.
With a third singlet right-handed neutrino it is also possible to have a
10 eV neutrino.
\newpage

\noindent {\bf 1. Introduction} --
The global and local symmetries encountered in particle physics have been
extremely useful in the study of particle physics. The difference between them
though, is profound and many physicists have argued that all the symmetries in
Nature must be gauge symmetries.
If we want to gauge a seemingly interesting global symmetry then, there are
some constraints that we have to consider. At low energies, most global
symmetries are broken and also we need to cancel all the anomalies.

In the case of a discrete symmetry we can entertain the idea
that the symmetry is a remnant from a spontaneously broken gauge or global
symmetry.
This may be the case if a continuous group $G$ is broken to a discrete
subgroup $H$. In addition, if the original group $G$ does not have an anomaly,
then it can be a discrete $Z_{N}$ gauge symmetry.
Recently there have been an interest in studying this possibility since
there is also an argument coming from quantum gravity \cite{qg}, that
wormhole effects should violate all global and discrete
symmetries.\footnote{Global symmetries may or may not have an anomaly. If it
does not have an anomaly, it can be gauged in principle.}
Discrete symmetries have played an important role in model building before,
and it is still useful to consider discrete gauge symmetries
\cite{w&k}, in the context of quantum gravity.

Supersymmetric extensions of the Standard Model is a good example
where there is a need to have discrete symmetries in order not to
let the proton decay too quickly. This has recently been reviewed in
\cite{i&r}
where the possibilities that the discrete symmetries are remnants from a
spontaneously broken $U(1)$ gauge symmetry are discussed.
Another area to deploy this idea is of course the Yukawa sector.
Still  very little is known about the structure of the mass matrices
in the Standard Model
and different attempts have been done to write down mass matrices
that gives the right phenomenology, notably the Fritzsch ansatz
\cite{fritzsch},
and the recent one by the authors of \cite{d&r&h}.
Most of these efforts have been devoted to the quark sector. Nowadays,
however,  we can as well contemplate neutrino
mass matrices since experimental
data have recently \cite{gallex} come about, that restrict this otherwise
$\lq\lq$many free
parameter" problem. One solution of the Solar neutrino problem,
the so--called MSW-mechanism \cite{msw}, is given by
particle physics theory and restricts the allowed regions in the mixing
parameter space of two massive neutrinos to be \cite{smirnov}
\be
m^{2}_{1}-m^{2}_{2} = (0.3-1.4)\times 10^{-5} {\rm eV}^{2}\hspace{6mm}
\sin^{2}2{\theta} = (0.4-1.3)\times 10^{-2}\\
\label{constr}\ee
\be
m^{2}_{1}-m^{2}_{2} = (0.35-9)\times 10^{-5} {\rm eV}^{2}\hspace{6mm}
\sin^{2}2{\theta} = 0.4-0.8
\label{constraints}\ee
Also we have upper limits on the masses of the three known neutrinos from
hadronic decays and limits on heavy neutrinos from LEP \cite{numass}.
Yet another experimental constraint on the effective Majorana mass comes from
the non-observation of neutrinoless double beta decay \cite{9}.

Here in this paper we are investigating different ways to introduce a
$U(1)$ gauge symmetry in the leptonic part of the Yukawa sector which is
subsequently spontaneously broken down to a $Z_{2}$ symmetry.
In general, for a $Z_{N}$ symmetry this is done by assigning the new
$U(1)$-charge for the Higgs field that breaks the symmetry by a VEV to be $Nq$
and for the rest of the fields to have charge quantized in units of $q$.
Throughout the paper we will assume that all the quarks carry zero charge.
We will show that the smallest extension of the Minimal Standard Model in
this scenario is a model with two extra right-handed singlet neutrinos and
two extra singlet Higgs fields.

\noindent{\bf 2. Models} --
There are essentially two different ways to pursue the idea of having
massive neutrinos included in the Standard Model \cite{b&p}:

1)  We can introduce Higgs triplets and write down mass terms with
only the left-handed neutrinos.

2)  We can introduce singlet
right-handed neutrinos and have singlet mass terms
including only the right-handed neutrinos. The singlet mass terms can also be
coupled to a Higgs singlet. When there is both left- and right-handed
neutrinos in the model there will also, in general, be Dirac mass terms.

We introduce the mass matrix for the neutrinos in a basis where the mass
matrix for the charged leptons is diagonal.
For simplicity, we will assume that the mass matrix
is real so that there is no CP-violation in the leptonic sector of our model.
Then the mass terms can be written as
\vskip0.5cm
\be
{\cal L}_{M}=-\frac{1}{2}{\bar N}\left( \begin{array}{cc}
M_{1}&D \\
D^{T}&M_{2}
\end{array} \right)N^{c}  + h.c
\label{matrix}\ee
\vskip0.5cm
where
\be
N=(\nu_{1L},..\nu_{jL},(\nu_{1R})^{c},..(\nu_{kR})^{c})
\label{basis}\ee
T means transpose and c means charge conjugation.
In general there is no need for $j$ and $k$ to be equal and we consider in this
paper $j=3$ and $k=1,2,3$.
$M_{1}$ refers to terms like $\overline{(\nu_{iL})^{c}}\nu_{jL}$,
$D$ refers to terms like ${\bar\nu_{iL}}\nu_{jR}$ and
$M_{2}$ refers to terms like $\overline{(\nu_{iR})^{c}}\nu_{jR}$.

At present there is no deep understanding about the sizes  of these entries,
if they are zero or non-zero, etc. In this paper we will investigate the
structure of this mass matrix imposed by a $Z_{2}$ symmetry that is a remnant
of a broken $U(1)$-symmetry. The anomalies of the original $U(1)$ symmetry need
to be cancelled and this fact will restrict the fermion content of the theory.
The various possibilities displayed in {\it Table 1}, which in some sense are
the smallest steps beyond the Minimal Standard Model, will be surveyed.
\begin{center}
Table 1 {\it The different charges of the extra $U(1)$ group.}
\end{center}
\begin{displaymath}
\begin{tabular}{|c|c|c|c|c|c|c|c|c|c|c|} \hline
$l_{e,L},e_{R}$&$l_{{\mu},L},\mu_{R}$&$l_{{\tau},L},\tau_{R}$&
$\nu_{1,R}$&$\nu_{2,R}$&$\nu_{3,R}
$&$H$&
$\Phi_{2}$&$\Phi_{3}$&$\Phi_{4}$&  \\ \hline
--1&1&0&-&-&-&0&-&2&-&case 1  \\ \hline
--1&1&0&0&-&-&0&0&2&-&case 2  \\ \hline
--1&1&0&0&-&-&0&0&-&2&case 3  \\ \hline
--1&1&0&--1&1&-&0&-&2&-&case 4  \\ \hline
--1&1&0&--1&1&-&0&-&-&--2&case 5  \\ \hline
--1&1&0&--1&1&-&0&0&2&-&case 6  \\ \hline
--1&1&0&--1&1&-&0&0&-&--2&case 7  \\ \hline
--1&1&0&--1&1&-&0&-&2&--2&case 8  \\ \hline
--1&1&0&--1&1&0&0&0&2&-&case 9  \\ \hline
\end{tabular}
\end{displaymath}
$l_{i,L}$, $e_{R}$, $\mu_{R}$ and $\tau_{R}$ are the three different lepton
doublets and the corresponding right-handed fields. $\nu_{i,R}$ are three
singlet right-handed neutrinos. $H\ (\equiv\Phi_{1})$ is the Higgs doublet
in the Minimal Standard Model. $\Phi_{2}$ is a Higgs singlet with zero charge
and with $<\Phi_{2}> \sim 10^{16}$ GeV.
$\Phi_{3,4}$ are Higgs singlets with $<\Phi_{3,4}> \sim 10^{16}$ GeV and
with $U(1)$ charges $\pm 2$.

In all the different cases at least one Higgs field with a non-zero charge
is needed in order to break the $U(1)$ symmetry.
The first case is severely restricted by LEP experiments. With only left-handed
neutrinos in the model we need a Higgs triplet in order to have mass terms.
A Higgs triplet, though,
would have contributed to the $Z$ decay width and is ruled out by LEP
\cite{higgstriplet}. Instead there is the
possibility to write down a non-renormalizable mass term, induced by gravity,
including two Higgs doublets.
\be
{\cal L}(\nu_{L},H)=
{\frac{\mu_{ij}}{M_{Pl}}}\overline{({\nu}_{iL})^{c}}\tau_{2}{\vec{\tau}}
{\nu}_{jL}\cdot {H}^{T}\tau_{2}{\vec{\tau}}{H}
\label{nonre}\ee
Here $\mu_{ij}$ are unknown coupling constants of order one and $\vec{\tau}$
are the Pauli matrices.
When the doublets acquire their vacuum expectation values this term will be
suppressed by typically a factor (250 GeV)$^{2}/10^{19}
{\rm GeV} \sim 10^{-5}$ eV.
This implies that we need some more scale in order to meet the requirements
from the MSW-solution of the Solar neutrino problem. It is important to notice
that this is a conclusion independent of the particular gauged models we have
in mind here and will of course not change if we introduce a
fourth family.  In the remainder of
this paper we will consider three families only.

So the other option is to introduce extra singlet right-handed neutrinos,
and at the same time disregard the non-renormalizable terms in the
model and the way to do it is restricted by the anomaly cancellation
requirements.  Since we assume that
none of the quarks carry the new charge, there are only three fermion doublets
left in the model that can carry charge and in order to cancel the
$SU(2)-SU(2)-U(1)$
anomaly two of the doublets carry opposite charges and the third doublet has
zero charge. This circumstance will prove to give some interesting
predictions that will be discussed later.

The first step, corresponding to cases 2 and 3, in this direction beyond the
Minimal Standard Model will then
be to introduce one extra singlet right-handed neutrino which carry charge
zero and also two singlet Higgs fields, $\Phi_{2}$ and one of $\Phi_{3}$ or
$\Phi_{4}$ with $U(1)$-charges 0,2,--2, in order to
have a new scale in our model and also break the $U(1)$ symmetry. The Higgs
singlets, will have $<\Phi_{2,3,4}> \sim 10^{16}$ GeV
and the mass matrix will take the form
\be
 \left( \begin{array}{cccc}
0&0&0&0 \\
0&0&0&0 \\
0&0&0&m_{4} \\
0&0&m_{4}&m_{5}
\end{array} \right)
\label{m4}\ee
$m_{4}$ are Dirac mass terms due to Minimal Standard Model Higgs doublet.
$m_{5}$ is a singlet mass term that is coupled to $\Phi_{2}$.
However, we cannot have $m^{2}_{1}-m^{2}_{2} \sim 10^{-5} {\rm eV}^{2}$
and at the same time have three light neutrinos as required by the
LEP-experiment. This is a conclusion independent on $<\Phi_{2,3}>$.
So case 2 and 3 do not work either.
By reconsidering the non-renormalizable
terms the two massless neutrinos will become massive but the previous problem
will still remain.

With two extra singlet right-handed neutrinos that carry charges --1 and 1
we will have the following mass matrix corresponding to cases 4--8
\be
 \left( \begin{array}{ccccc}
0&0&0&m_{3}&0 \\
0&0&0&0&m_{4} \\
0&0&0&0&0 \\
m_{3}&0&0&m_{5}&m_{6} \\
0&m_{4}&0&m_{6}&m_{7}
\end{array} \right)
\label{m5}\ee
$m_{3}$ and $m_{4}$ are Dirac mass terms from the Standard Model Higgs
doublet. $m_{5}$ and $m_{7}$ are couplings to Higgs singlets that carry
charges +2 and --2 respectively. $m_{6}$ is a coupling to a Higgs singlet that
carry zero charge.
All the $m_{i}$ need not be non-zero. Our aim is to have the minimal version
that gives the right masses and mixings without using, too much, the freedom
in assigning strengths to the different Yukawa coupling constants.
It turns out that in order to get the right mass eigenvalues two of the
$m_{i}$; $i$=5,6,7 need to be non-zero and the
$<\Phi_{3,4,5}> \sim 10^{16}$ GeV.
However, if we let $m_{6}$=0 then the mass matrix will become block diagonal
in such a way that the two light neutrinos will not mix, which of course is
not acceptable to us.
So the first model beyond the Minimal Standard Model where we can have the
correct phenomenology would then correspond to case 6 or 7 with the mass
matrix  given in (\ref{m5}) where $m_{5}$=0 or $m_{7}$=0.
For definiteness, let us choose $m_{7}$=0.
Here we must
use the freedom in the Yukawa strength of the $m_{3}$ and $m_{4}$ entries.
For the light neutrinos the eigenvalue equation is an approximate second order
equation and the solution is the following
\be
{\lambda}_{1,2}=\frac{\beta^{2}{m_{4}}^{2}}{2m_{6}}
(1\pm\sqrt{1+4\alpha^{2}\beta^{-2}})
\label{solution}\ee
where $\alpha \equiv m_{3}/(250$ GeV) and $\beta \equiv m_{4}/(250$ GeV)
in (\ref{m5}) and $m_{5}=m_{6}=10^{16}$ GeV.
We can easily meet the requirements of (\ref{constr}) or (\ref{constraints});
\be
\alpha = 0.036,\hspace{2mm} \beta = 0.71\ \
\Rightarrow \sin^{2}2\theta=10^{-2},\hspace{6mm} {\lambda}_{1}^{2}-
{\lambda}_{2}^{2}=1.0\times 10^{-5}{\rm eV}^{2}
\label{0.036}\ee
\be
\alpha = 0.39, \hspace{2mm} \beta=0.63\ \
\Rightarrow \sin^{2}2\theta=0.6,\hspace{6mm} {\lambda}_{1}^{2}-
{\lambda}_{2}^{2}=1.0\times 10^{-5}{\rm eV}^{2}
\label{0.74}\ee
In the first case the masses are $3.2\times 10^{-3}$ eV and
$ 2.5\times 10^{-5}$ eV
whereas in the second case they are  $3.2\times 10^{-3}$ eV and
$ 0.73\times 10^{-3}$ eV.
In both cases there are also one massless neutrino, that does not mix with
the other neutrinos, and two neutrinos with
masses of order $10^{16}$ GeV. The massless neutrino could be massive if we
reconsider the possibilities to have effective non-renormalizable mass terms
corresponding the $M_{1}$-part of (\ref{matrix}). However, it will still be
decoupled and the somewhat unorthodox scenario where the electron neutrino
is not the lightest one will emerge. This is due to the fact we are considering
only lepton doublets carrying charge and in order to cancel the anomaly two
of the doublets have opposite charges and the third has zero charge. This makes
the matrix in (\ref{m5}) block diagonal.

There is another way to make the massless neutrino become massive.
Let us introduce a third extra right-handed neutrino with zero charge,
corresponding to case 9 in {\it Table 1}.
This will give us the following mass matrix
\vskip0.5cm
\be
 \left( \begin{array}{cccccc}
0&0&0&m_{3}&0&0 \\
0&0&0&0&m_{4}&0 \\
0&0&0&0&0&m_{8} \\
m_{3}&0&0&m_{5}&m_{6}&0 \\
0&m_{4}&0&m_{6}&m_{7}&0\\
0&0&m_{8}&0&0&m_{9}
\end{array} \right)
\label{m6}\ee
\vskip0.5cm
The matrix is block diagonal and the submatrix with the entries $0, m_{8}$
and $m_{9}$ will give us two massive neutrinos. Remember that $m_{8}$ is due
to the ordinary Higgs doublet and that $m_{9}$ is due to a Higgs singlet
with zero charge and with a vev of $10^{16}$ GeV. Here in this case we can
for example put $m_{8}\sim250$ GeV and $m_{9}\sim10^{13}$ GeV
and this gives us
two massive neutrinos of order 10 eV and  $10^{16}$ GeV.
This last case has the nice feature that the electron neutrino can be
identified as the lightest one and that we also have a dark matter candidate
in the spectrum, a 10 eV tau neutrino \cite{dm} that does not mix with the
electron neutrino or the muon neutrino.

\noindent {\bf 3. Conclusions} --
Here in this paper we have investigated the structure imposed by discrete
gauge symmetries on the neutrino mass matrix in models slightly beyond the
Minimal Standard Model. The discrete gauge symmetry we study is a $Z_{2}$
subgroup of an extra $U(1)$ gauge symmetry where only scalars and leptons
carry charge.
Models with no or one extra singlet right-handed
neutrino are ruled out.
With two extra singlet right-handed neutrinos and two extra
singlet Higgs field we can construct a model that fits data. The model predict
one massless neutrino and that the electron neutrino is not the lightest one.
With three extra singlet right-handed neutrinos and two
singlet Higgs field we can construct a model where the electron neutrino is
the lightest one and where there is also 10 eV tau neutrino in the spectrum,
a Dark Matter candidate, that does not mix with the electron neutrino or the
muon neutrino.
{\bf \center Acknowledgements \\}
B-{\AA} Lindholm is supported by the Center for Theoretical Physics,
Seoul National University and by The Swedish Institute.
\bibliographystyle{unsrt}

\end{document}